\documentclass[a4paper]{article}
\usepackage{amsmath,graphicx}
\usepackage{graphicx}
\usepackage{fullpage}
\usepackage[utf8]{inputenc}
\usepackage{lipsum}  
\usepackage{amssymb,amsmath,array}
\usepackage{amsthm}
\usepackage[hidelinks]{hyperref}
\usepackage{xspace}
\usepackage{bbm}
\usepackage{amsmath}
\usepackage{amsthm}
\usepackage{amssymb}
\usepackage{graphicx} 
\usepackage{manfnt}
\usepackage{url}
\usepackage{xcolor}
\usepackage{soul}
\usepackage{amsmath}
\usepackage[british]{babel}
\usepackage{wrapfig}
\usepackage{subfigure}
\usepackage[font=small]{caption}

\DeclareMathOperator*{\sign}{sign}
\newcommand{\scp}[3][]{#1\langle #2, #3 #1\rangle}
\newcommand{\rop}{\cl A^{\rm v}}
\newcommand{\ropmtx}{\cl A}
\newcommand{\drop}{\cl B}
\newcommand{\ie}{\emph{i.e.}, }
\newcommand{\eg}{\emph{e.g.}, }
\newcommand{\sopu}{{\scriptscriptstyle\rm opu}}

\newcommand{\bs}{\boldsymbol}
\newcommand{\bb}{\mathbb}
\newcommand{\cl}{\mathcal}
\newcommand{\ts}{\textstyle}

\newcommand{\iid}{%
	\ifmmode
	\mathrm{i.i.d.}%
	\else%
	i.i.d.\@\xspace%
	\fi%
}

\newcommand{\edt}[1]{{\color{black} #1}}

\newcommand{\ignore}[1]{}

\newtheorem{theorem}{Theorem}[section]

\title{Signal processing with optical quadratic random sketches\vspace{-2mm}}
\author{Rémi Delogne$^*$, Vincent Schellekens$^\dagger$, Laurent Daudet$^\top$, Laurent Jacques$^*$\thanks{Part of this research was supported by the Fonds de la Recherche Scientifique – FNRS under Grant T.0136.20 (Project Learn2Sense).}\\[2mm]
	\small $^*$ICTEAM, UCLouvain, Belgium. $^\dagger$CEA, Paris. $^\top$LightOn, Paris}

\begin{document}
	%
	\maketitle
	\begin{abstract}
		Random data sketching (or projection) is now a classical technique  enabling, for instance, approximate numerical linear algebra and machine learning algorithms with reduced computational complexity and memory. In this context, the possibility of performing data processing (such as pattern detection or classification) directly in the sketched domain without accessing the original data was previously achieved for linear random sketching methods and compressive sensing. In this work, we show how to estimate simple signal processing tasks (such as deducing local variations in a image) directly using random \emph{quadratic} projections achieved by an optical processing unit. The same approach allows for naive data classification methods directly operated in the sketched domain. We report several experiments confirming the power of our approach.  
	\end{abstract}
	\section{Introduction}
	\label{sec:intro}
	An ever increasing number of algorithms in the area of machine learning, signal processing, and numerical linear algebra leverage \textit{random data sketching} (or \emph{random projections}) techniques to reduce the dimension of input data (or their number) and alleviate computational complexity up to a controlled approximation error \cite{achlioptas_database-friendly_2001,rahimi_random_nodate,baraniuk_simple_2008}. These methods are data-agnostic while preserving essential information about the instances they transform. In a similar way, the same techniques can also be used to lift a given signal or pattern to a higher dimensional space where data might be easier to classify (similar to kernels for example) \cite{saade_random_2016}. 
	
	At the output of the data sketching, one will often require to estimate specific properties of the embedded information (such as signals or images). One can think for example of a video stream in traffic monitoring or industrial quality control where only a part of the video frames will later require attention, calling for means of restricting the frames, after sketching, to specific area of interest \cite{ka03}. 
	
	Such estimation can easily be achieved when the essence of the sketching is linear \cite{SPWCM}. In this work however, we focus on the non-linear sketching mechanism achieved by an optical processing unit (OPU) that allows super fast \edt{and ultra-low energy} computing of random data sketching in the optical domain \cite{saade_random_2016,ohana_kernel_2020}. As will be made clear in Sec.~\ref{sec:opu}, this device is able to reliably compute in parallel millions of \textit{quadratic random projections} of signals or images. Moreover, these projections are tantamount to applying \textit{rank-one projection} (ROP) of a lifted version of the input \cite{chen_exact_2015,cai_rop_2015} where the implicit random rank-one matrix is hardly accessible (if not completely inaccessible) to us (Sec.~\ref{sec:sketching}). \edt{Ultimately, having access to such a powerful yet energy efficient projection tool could lead to applications to processing of large data streams such as videos using very little power.}

	In our previous work (\cite{esann22}), we showed theoretically that up to some controlled distortion, we can operate linear signal estimation directly in the sketched domain without ever reconstructing the original signal, thus avoiding costly reconstruction methods \cite{chen_exact_2015,cai_rop_2015,phaselift,foucart_flavors}. We illustrated this using simulations of the ROP on classical computers, in the hope of getting a foretaste of what the OPU is capable of. In this work we show how the OPU is capable to put these theoretical results in practice. Though straightforward in theory, the OPU requires some fine-tuning to be used in practice as we shall see in further sections.
	
	\section{Sketching with hidden rank one projections}
	\label{sec:sketching}
	This section presents the mathematical framework of \textit{quadratic sketching}. As described in the introduction, this transformation relies on the \textit{rank-one projection} (ROP) of the observed signal and it is indeed crucial to understand this before getting to the description of the OPU.
	
	{Quadratic random sketching} consists in taking a series of $m$ measurements $(\bs a_i^\top\bs x)^2$ of a signal of interest $\bs x\in\bb R^n$, with a set of $m$ random vectors $\{\bs a_i\}_{i=1}^m \subset \bb R^n$. The sketching operator $\rop$ is defined as
	\begin{equation}
		\label{eq:originalROP}
		\rop: \bs x \in \bb R^n \mapsto \rop(\bs x) := \big( (\bs a_i^\top\bs x)^2 \big)_{i=1}^m \in \bb R^m_+.
	\end{equation}
	
	\ignore{As discussed in the previous section, we assume that while we can compute the operator $\rop$, we cannot access the sketching vector $\{\bs a_i\}_{i=1}^m$. The OPU allows us to compute all the components of $\rop(\bs x)$ in a reproducible way using the physical properties of multiple scattering of coherent light in random media, which is thus extremely fast and power-efficient (even if $m \simeq n$). In this context, the vectors $\{\bs a_i\}_{i=1}^m$ are fixed, but \emph{hidden} to us. Moreover, following the observations made in~\cite{saade_random_2016}, we assume that each random vector $\bs a_i$ is \iid as a Gaussian random vector $\bs a \sim \cl N(\bs 0, \bs I_n)$, with identity covariance $\bs I_n$. This optical projection is actually modelled by quadratic projections over complex random vectors\footnote{\ie with $a_i=a+bi$, with $a\sim\cl N(0,\frac{1}{2})$ and $b\sim\cl N(0,\frac{1}{2})$}, but we here work in the real field for the sake of simplicity.}
	
	As observed in the context of phase retrieval~\cite{phaselift}, the operator $\rop$ amounts to a ROP of the \emph{lifted signal}, \ie the rank-one matrix $\bs X = \bs x \bs x^\top \in \bb R^{n \times n}$, onto the rank-one random matrices $\{\bs A_i := \bs a_i \bs a_i^\top\}_{i=1}^m \subset \bb R^{n \times n}$, as defined by the equivalence $\rop(\bs x)=\big( (\bs a_i^\top\bs x)^2 \big)_{i=1}^m=\big( \bs a_i^\top\bs x\bs x^\top\bs a_i \big)_{i=1}^m=\big( \langle\bs A_i,\bs X\rangle \big)_{i=1}^m =: \ropmtx(\bs X)$, where $\langle \cdot, \cdot \rangle$ is the Frobenius inner product~\cite{chen_exact_2015,cai_rop_2015}. We thus use ``quadratic sketch'' and ``ROP measurements'' interchangeably.   
	
	As explained in Sec.~\ref{sec:opu}, the OPU allows us to compute all the components of $\rop(\bs x)$ in a reproducible way using the physical properties of multiple scattering of coherent light in random media, which is thus extremely fast and power-efficient (even if $m \simeq n$). In this context, the vectors $\{\bs a_i\}_{i=1}^m$ are fixed, but \emph{hidden} to us. Moreover, an asymptotic analysis made in~\cite{saade_random_2016} shows that each random vector $\bs a_i$ are very close to be identically and independently distributed (\iid) as a Gaussian random vector $\bs a \sim \cl N(\bs 0, \bs I_n)$, with identity covariance $\bs I_n$. 
	
	The ROP operator is sadly \textit{biased} (non-isotropic). In other words, the expectation of the squared norm of the ROP of a vector $\bs x$ is not proportional to the squared Frobenius norm of the lifted signal $\bs x\bs x^\top$ (equal to the fourth power of the norm of $\bs x$) \cite{chen_exact_2015}. \edt{This is relatively obvious considering that the ROP operator only outputs positive values.} However, isotropy is a key property to ensure that the ROP sketch keeps essential information on $\bs x$. This leads to the definition of a debiased ROP operator (DROP) which rids us of the bias (see \cite{chen_exact_2015}, lemma 4 and appendix F):
	\begin{equation}
		\label{eq:DROP}
		\drop: \bs x \in \bb R^{n} \mapsto \drop(\bs x)= \big( \rop_{2i}(\bs x) - \rop_{2i+1}(\bs x) \big)_{i=1}^{m}.
	\end{equation}
	
	This new debiased estimator DROP can easily be implemented on an OPU by first applying the operator $\rop$ on a vector then splitting it in two and subtracting one half from the other. Thanks to the constant computational complexity of $\rop$ on the OPU, this has little impact on the total computing time. 
	
	\section{Signal processing in the sketched domain}
	\label{sec:estim-sketch}
	

	As shown in \cite{SPWCM}, for linear random sketching technique $\bs x \mapsto \bs A \bs x$,  \edt{one can estimate $\scp{\bs u}{\bs x}$} as soon as the $m \times n$ matrix $\bs A$ satisfies the restricted isometry property (RIP) over $k$-sparse (or low-complexity) signals, \ie for any $k$-sparse $\bs x$ and $\bs u$, 
	
	$$
	|\scp{\bs A \bs u}{\bs A \bs x} - \scp{\bs u}{\bs x}| \leq \delta \|\bs u\|\|\bs x\|,
	$$
	for some small, controlled distortion $\delta$.  
	
	\edt{In our case we aim instead to recover $|\scp{\bs u}{\bs x}|^2$} thanks to a useful tool called the Sign Product Embedding (SPE) \cite{esann22,jacques_spe}. The SPE states that with high probability, $\scp{\sign(\drop(\bs u))}{\drop(\bs x)}$ acts as a proxy for $|\scp{\bs u}{\bs x}|^2$. A wise choice of $\bs u$ could hence give us access to local information about $\bs x$ for example, using only the sketch $\drop (\bs x)$ and $\drop(\bs u)$ without ever reconstructing $\bs x$. This is more formally stated in the following theorem.
	
	\begin{theorem}[Sign Product
		Embedding of DROP Sketches]
		\label{prop:drop-spe}
		Given a fixed unit vector $\bs u \in \bb R^n$, $\kappa = \pi/4$, and a distortion $0<\delta <1$, provided that 
		\begin{equation}
			\label{eq:sample-complex-drop-spe}
			\ts \ts m \geq C \delta^{-2} k \log(\frac{n}{k\delta}),    
		\end{equation}
		then, with probability exceeding $1 - C \exp(-c \delta ^2 m)$, for all $k$-sparse signals $\bs x \in \Sigma_k := \{\bs v \in \bb R^n: |{\rm supp}(\bs v)|\leq k\}$, $\cl B$ respects the SPE over $\Sigma_k$, \ie 
		\begin{equation}
			\label{eq:drop-spe}
			\ts \Big|\frac{\kappa}{m} \scp{\sign(\drop(\bs u))}{\drop(\bs x)} - {\scp{\bs u}{\bs x}^2}\ignore{{\|\bs u\|^2}} \Big| \leq \delta \|\bs x\|^2,    
		\end{equation}    
		with $\sign$ the sign operator applied componentwise on vectors. 
	\end{theorem}
	
	As detailed in \cite{esann22}, the proof of this key result consists of three main steps. The first shows that $\frac{\kappa}{m}\bb E\scp{\sign(\drop(\bs u))}{\drop(\bs x)}$ $=\scp{\bs u}{\bs x}^2$ using the rotational invariance of the Gaussian distribution that characterises $\drop$. As a second step, we can prove that for a fixed unit vector $\bs u$, and for a vector $\bs x\in\Sigma_k$ the random variable $\scp{\sign(\drop(\bs u))}{\drop(\bs x)}$ concentrates around its mean, using the properties of \textit{sub-Gaussian} random variables. In the final step, the previous result is extended by continuity on all unit vectors $\bs x\in\Sigma_k$. For this purpose we use a union bound over a covering of the lifted space of rank-one matrices and bound the deviation between $\scp{\sign(\drop(\bs u))}{\drop(\bs x)}$ and $\scp{\bs u}{\bs x}^2$ over the balls of the covering. Combining the union bound over the covering and the probability of failure over a ball of the covering yields the desired result.
	\medskip 
	
	In words, this theorem simply shows that if $m$ is large enough and if the DROP is defined by \iid random Gaussian vectors, the quantity $\frac{\kappa}{m} \scp{\sign(\drop(\bs u))}{\drop(\bs x)}$ approximates $\scp{\bs u}{\bs x}^2$ with a controlled distortion $\delta$ scaling like $O\big(\sqrt{k/m}\,\big)$ (up to log factors). It is therefore possible to approximate linear functions of any signal $\bs x\in\bb R^n$, by projecting their sketches $\drop(\bs x)$ on $\sign(\cl B(\bs u)) \in \{\pm 1\}^m$. 
	
	\section{The optical processing unit}
	\label{sec:opu}
	Armed with all the necessary tools and information, we can now dive into the working principles of the OPU. As we touched upon in the introduction, the aforementioned sketching operators ($\rop, \drop$) are motivated by the fact that they can be calculated by an OPU with the same computational complexity regardless of both the dimension of the signal we want to project and the sketch dimension (up to maximal dimensions determined by the optical setup). This section details essential aspects of the OPU.
	
	\begin{figure}[t]
		\centering
		\includegraphics[width=8.5cm]{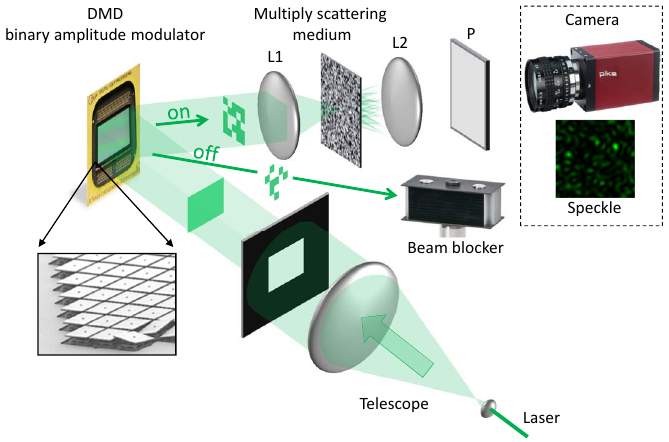}
		\caption{\cite{saade_random_2016} The OPU takes a laser beam and uses an array of mirrors to encode binary data in the beam \cite{liutkus2014imaging}. It then sends the beam through a scattering medium replicating the effect of the operator $\rop$ on the signal encoded in the light beam.}\medskip
		\label{fig:opu}
	\end{figure}

	As summarised in Fig.~\ref{fig:opu}, the OPU processes a binary input\footnote{Recent extensions of the OPU admit non-binary $x$ thanks to a bit-plane representation, or by macropixel encoding \cite{saade_random_2016}.} $\bs x \in \{0,1\}^n$ (with $n$ up to $O(10^6)$) by using it to program the orientation of a grid of mirrors in a digital micro-mirror device (DMD). In this DMD, only correctly oriented ``on'' (1) mirrors (as opposed to mis-aligned ``off'' (0) mirrors) can reflect part of an incident, coherent laser beam to a stable scattering medium. After scattering, the intensity of a complex pattern, or \emph{speckle}, arising from constructive and destructive interferences of the scattered light, is recorded on the focal plane of a camera.  
	
	By correctly adjusting the pixel pitch of the camera to the correlation length of that speckle, it was shown in \cite{saade_random_2016} that the so-called transmission matrix of the medium is very close to a fixed random matrix with Gaussian i.i.d. entries. Therefore, each camera pixel records independent measurements
	\begin{equation}
		\label{eq:opu-pure-sensing}
		y \sim \gamma |\scp{\bs a}{\bs x}|^2,\text{ with } \bs a \in \bb R^n,\ a_i \sim_{\iid} \cl N(0,1),
	\end{equation}
	where $\gamma>0$ is a certain conversion gain. 
	
	In practice, this idealised model undergoes a few alterations. First, the camera records intensity values uniformly coded over 8 bits of information (ranging over integers from 0 to $255=2^8-1$), with a saturation (or \emph{clipping}) level $C>0$ met (by design) by less than 1\% of the pixels. Second, the sensing model is corrupted by a prequantisation noise (mainly due to electronic noise, photon counting, and scaterring medium variations) whose amplitude mainly impacts the two first encoding bits. We note that other less important effects (not considered here for simplicity), such as focal plane vignetting, also impact the OPU output. The actual OPU data transformation is thus closer to the model
	\begin{equation}
		\label{eq:opu-noisy-sensing}
		y = \cl Q_b(\gamma |\scp{\bs a}{\bs x}|^2 + \edt \eta),
	\end{equation}
	with the uniform quantiser $\cl Q_b(t)$ equal to $\delta \lfloor t/\delta \rfloor$ if $0\leq t\leq C$, and $C$ otherwise, the bin width $\delta = C 2^{-b}$, the bit depth $b=8$, and a noise $\edt \eta$ such that $|\edt \eta|\leq 4 \delta$ with high probability.    
	
	Therefore, when the OPU records an $m$-length measurement vector   
	$\bs y_{\sopu} = \rop_{\sopu}(\bs x) \simeq \cl Q_b(\rop(\bs x) + \bs{ \edt \eta})$, with a noise vector $\bs{\edt\eta} =(\edt \eta_1, \ldots, \edt \eta_m)^\top$, and $\rop$ encoding $m$ \iid Gaussian random vectors $\{\bs a_j\}_{j=1}^m$, both the quantisation and the prequantisation noise alter the pure ROP sketch $\rop(\bs x)$, as well as the DROP $\drop_{\edt{\sopu}}(\bs x)$ computed from it. In particular, we can expect severe alterations when the square norm of the input vector $\bs x$, \ie its number of ones, is either too small (in this case $\bs y_{\sopu}$ is dominated by the noise variations) or too large, in which case many camera pixels will saturate. 
	
\begin{figure}[t]
		\centering
		\includegraphics[width=\columnwidth]{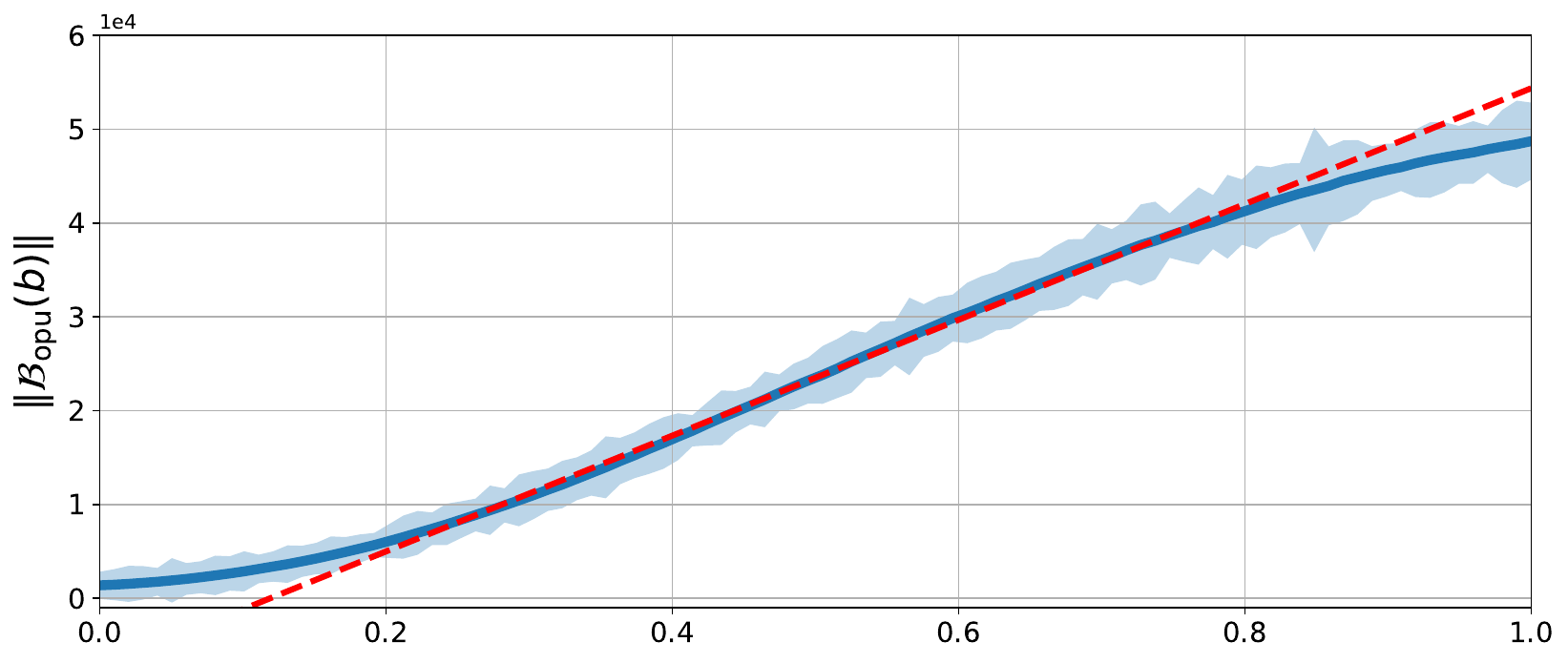}
		\caption{Evolution of $\|\drop_{\edt{\sopu}}(\bs b)\|$ vs. $s=\|\bs b\|^2$, for various values of $s$ (in percent of $n$). The blue area has local thickness of $100\times$std.}
		\label{fig:opu-alterations}
	\end{figure}

	We evaluate this effect in Fig.~\ref{fig:opu-alterations}. We have there represented $\|\drop_{\sopu}(\bs b)\|$ by randomly sampling binary vectors $\bs b\in\bb R^{n=10^6}$, with sparsity level $s := \|\bs b\|^2 = |{\rm supp}(\bs b)|$ ranging from 0 to $10^6$. The quantity $\drop_{\sopu}(\bs b)$ was computed from $\rop_{\sopu}(\bs b)$ with $m=10^4$ measurements. For each value of $s$, the OPU was run 20 times to study the impact of noise. In absence of disturbance to the model, we \edt{should} have $\|\rop(\bs b)\|$, and thus $\|\drop(\bs b)\|$, proportional to $s$. However, Fig.~\ref{fig:opu-alterations} shows this linear relation only holds approximately for $s$ between 20\% and 80\% of $n$. Before this range, the noise dominate and leads to meaningless observations, and after it, an increasing number of pixels saturate which decreases the value of $\|\drop_{\sopu}(\bs b)\|$ compared to $\|\drop(\bs b)\|$. Moreover, in the linear regime, the variations of $\|\drop_{\sopu}(\bs b)\|$ between different OPU calls show that $s$ must sufficiently increase for $\|\drop_{\sopu}(\bs b)\|$ to exceed noise variation (\eg $\Delta s = 500$ from a naive error analysis of Fig.~\ref{fig:opu-alterations}). \edt{This experiment shows that the OPU can convincingly be used to approximate $\drop$ provided we respect a certain level of sparsity guaranteeing good behaviour.}

	\section{Experiments}
	
	\edt{With theoretical guarantees as well as experimental validation of the OPU, we present in this section two experiments demonstrating the possibility of performing basic signal processing and classification tasks in the sketched domain provided by an OPU}.
	
\begin{figure}[t]
		\centering
		\raisebox{5mm}{\includegraphics[width=.34\columnwidth]{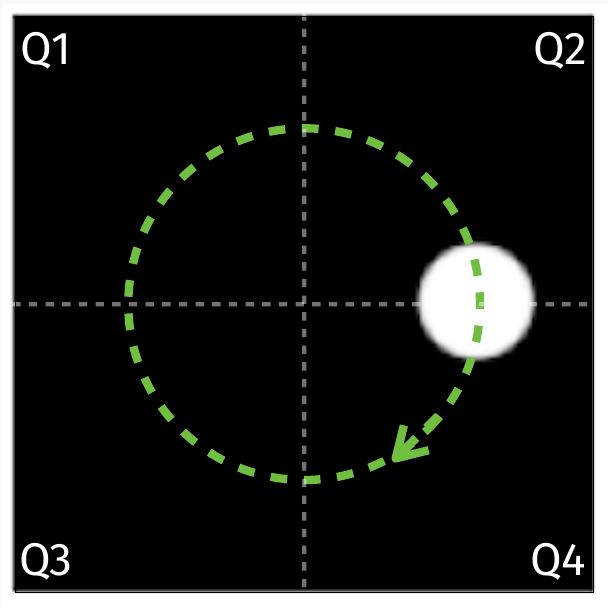}}
		\includegraphics[width=.64\columnwidth]{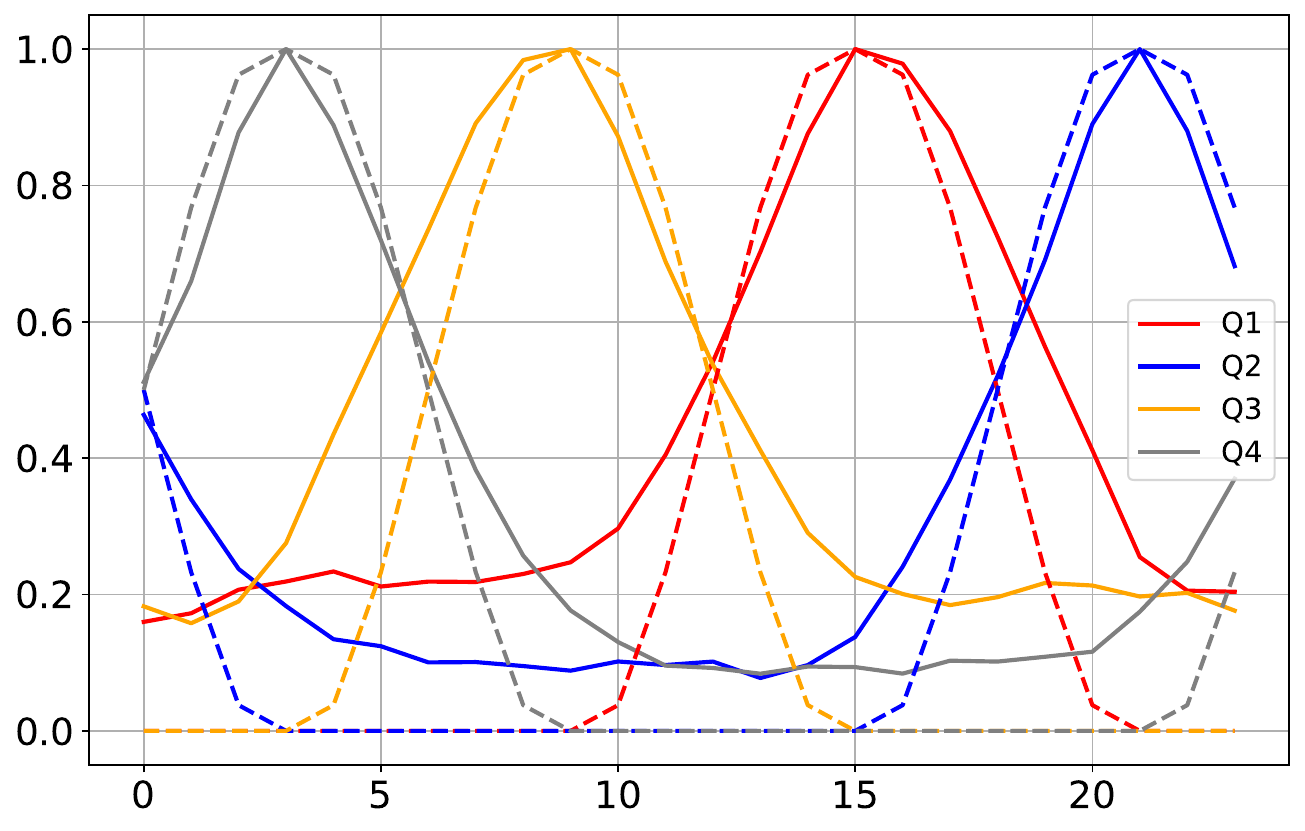}
		\caption{(Left) A frame from the synthetic video of a white rotating disk on a black background. (Right, best seen in colour) Evolution of the (normalised) true and estimated quadrant occupancies, in dashed and plain curves, respectively, for each quadrant in function of $0\leq t \leq 23$ according to the colour scheme detailed in the legend.}
		\label{fig:disk}
	\end{figure}

	As a first experiment, we consider a synthetic video consisting of 24 vectorised $950\times 950$ binary images $\{\bs x_t\}_{t=0}^{23}$ (\ie vectors of dimension $n=902\,500$), representing a (white) rotating disk on a black (zero) background (see Fig.~\ref{fig:disk}(left)). Following Sec.~\ref{sec:estim-sketch}, our objective is to detect the passage of the disk in each of the four quadrants of the image solely based on the $m-$dimensional OPU measurements $\{\drop_{\sopu}(\bs x_t)\}_{t=0}^{23}$. 
	
	We thus define four quadrant indicators, \ie four vectors $\bs u_i\in\bb R^n$ equal to 1 in the $j$-th quadrant and zero outside ($j\in\{1,\ldots,4\}$). Thanks to Thm~\ref{prop:drop-spe}, we can estimate, up to some distortion, the $j$-th \emph{quadrant occupancy} signal $q_j(t) := |\scp{\bs u_j}{\bs x_t}|^2$---the square of the integration of $\bs x_t$ in the $j$-th quadrant---from the estimated occupancy 
	\begin{multline*}
		\ts q^{\rm est}_{j,\sopu}(t) := \frac{\kappa}{m} \scp{\sign(\drop_{\sopu}(\bs u))}{\drop_{\sopu}(\bs x_t)}\\ 
		\ts \simeq q^{\rm est}_j(t) := \frac{\kappa}{m} \scp{\sign(\drop(\bs u))}{\drop(\bs x_t)}.
	\end{multline*}
	
	The curves in Fig.~\ref{fig:disk}(right) show the evolution with time of $q^{\rm est}_{j,\sopu}(t)$ with $m=10\,000$, compared to the true quadrant occupancies $q_j(t)=|\scp{\bs u_j}{\bs x_t}|^2$. Each curve has been normalised to its maximum value to attenuate their dependence in the (unknown) gain factor $\gamma$ in the model \eqref{eq:opu-noisy-sensing}. The comparison of these curves shows that information about the quadrant occupancy of the disk is preserved in the quadratic OPU sketches, thanks to the estimations of $q^{\rm est}_{j,\sopu}(t)$. 
	
	\medskip
	Our second experiment consists in a simple classification task on a binarised version of the handwritten digits of the MNIST dataset. This dataset $\cl X := \{(\bs x_k, t_k)\}_{k=1}^{N=70\,000} \subset \{0,1\}^{n=28^2} \times \{0,\ldots,9\}$~\cite{mnist} is obtained by a mere thresholding (at threshold $2^6$) of each 8-bit graylevel image of MNIST.   
	
	We here aim to compare the result of a naive classification performed in the direct domain, to a classification operated in the sketched domain. We start by randomly splitting $\cl X$ into a training $\cl X_{\rm tr}$ and a test set $\cl X_{\rm te}$ according to a split of $60\,000$ and $10\,000$ images, respectively. From $\cl X_{\rm tr}$ we compute the 10 centroids $\{\bs c_j\}_{j=0}^9 \subset \bb R^{n}$ of each class of digits and we define 10 vectors $\bs u_j = \bs c_j /\|\bs c_j\|$. In the direct domain and for images of the test set, the estimated label of an image $\bs x_k$ is then defined as $\hat{t}_k := \arg\max_j |\scp{\bs u_j}{\bs x_k}|^2$. 
	
	From the observations made in Sec.~\ref{sec:opu}, directly sketching the MNIST instances (\eg by zero-padding them at the right dimension before injection in the OPU) poses a particular challenge as each $28\times 28$ binarised image (with $n=784$) may not contains enough ones to reach a high signal-to-noise ratio at the OPU output. This is solved by expanding each binary MNIST image $\bs x$, seen as a matrix $28\times 28$ binary matrix, with a simple Kronecker product $\cl I(\bs x) := \bs x \otimes \bs M$ with a $p\times p$ matrix $\bs M$ of ones, \ie each pixel of $\bs x$ is turned in a $p\times p$ macro pixel, and the inflated image $\cl I(\bs x)$ gets a dimension $n=784 p^2$ that nears the SLM resolution, with $p^2$ more ones than in the initial images. 
	
	After sketching these inflated instances, the label estimate of a test image $\bs x_k$ is computed by comparing the two transformations 
	$$
	\ts \hat{t}^{\rm sk}_k := \arg\max_j \frac{\kappa}{m} \scp{\sign(\drop_{\edt\sopu}(\bs u_j))}{\drop_{\edt\sopu}(\bs x_k)},
	$$
	which should be close to $\hat{t}_k$ according to Thm~\ref{prop:drop-spe}. We summarise in Table~\ref{tab:MNIST-table} the different average testing accuracies reached for both the direct and the sketched classifications. Keeping in mind that our classification method is rudimentary (as shown in, \eg \cite{saade_random_2016}, it is possible to develop much better, non-linear classification algorithms directly in the sketched domain), we observe anyway that as $m$ increases the average accuracy improves and approaches the one performed in the direct domain. The loss in accuracy can also be attributed to several other factors such as the binarisation process and noise within the OPU.  
	
	\begin{table}[t]
		\centering
		\scriptsize
		\begin{tabular}{c|cccccc}
			& Direct & $m=200$ 
			& $400$ & $800$ & $1600$&$3200$
			\\ 
			&&&&&&\\[-2.5mm]
			\hline
			&&&&&&\\[-2mm]
			Accuracy  $[\%]$ & $82.1$ & $56.3$
			& $59.2$ & $66.9$&$71.8$&$75.0$
		\end{tabular}
		\caption{Testing accuracy (in $\%$) in the direct domain (without sketching), and in the sketched domain for various values of $m$.}
		\label{tab:MNIST-table}
	\end{table}
	\medskip
	
	Interestingly, the SPE described in \eqref{eq:drop-spe} allows us to devise another classification technique, directly trained in the sketched domain, and which outperforms the previous method. For this, we consider the sketched dataset $\cl X^{\drop_{\edt\sopu}}= \{(\drop_{\sopu}(\bs x_k), t_k)\}_{k=1}^{N=70\,000}$, and compute all the centroids $\{\bs c^{\drop_{\edt\sopu}}_{j}\}_{j=0}^9 \subset \bb R^m$ in the training split $\cl X^{\drop_{\edt\sopu}}_{\rm tr}$ of $\cl X^{\drop_{\edt\sopu}}$. 
	
	By assuming that, for every such centroid, there exists a vector $\bs v_j \in \bb R^n$ such that $\bs c^{\drop}_j \approx \drop(\bs v_j)$, we can estimate our label by applying a \emph{sign} operation to each $\bs c^{\drop}_j$---which is aligned to Thm~\ref{prop:drop-spe}---and compute 
	$$
	\ts \hat{t}^{\drop, \sign}_k := \arg\max_j \scp{\sign(\bs c^{\drop_{\edt \sopu}}_j)}{\drop_{\edt\sopu}(\bs x_k)}.
	$$
	With $m=1\,000$ we now reach an accuracy of $82.7\%$ (and $83.9\%$ at $m=10\,000$), slightly better than the classification accuracy achieved in the direct domain of MNIST.
	
	\section{Conclusion}
	
	In this work we built upon the theoretical foundations laid in \cite{esann22} in order to show that signal estimation in the sketched domain is possible with the optical quadratic random sketching delivered by an OPU. We reminded the reader of the mathematical tools (such as the biased and debiased rank-one projections, and the sign product embedding) and analysed the working principle of the OPU as well as its deviation to the pure quadratic sensing model. We then applied our proposed method to two toy examples to demonstrate the possibility of extracting localised information from a sketched signal and classifying images from their sketches.
	
	Future works could exploit both the OPU calibration (as studied in \ref{sec:opu}) and the dependence of the SPE distortion in the sketch dimension to formalise precise statistical tests for pattern matching in the sketched domain. On a more theoretical note, preliminary numerical tests show that the SPE of the DROP could hold for matrices of rank greater than 1. This is appealing for, \eg change point detection application in a data stream $\{\bs x_t\}_{t \in \bb Z}$ since, for instance, a time change in ROP sketches $\rop(\bs x_{t+1}) - \rop(\bs x_{t})$ is equivalent to the ROP of the rank-2 matrix $\bs x_{t+1}\bs x_{t+1}^\top - \bs x_{t}\bs x_{t}^\top$.

\end{document}